\documentclass[AMA,STIX1COL]{WileyNJD-v2}

\usepackage{booktabs}
\usepackage{graphicx}
\graphicspath{{figures/}}
\usepackage{enumitem}
\usepackage{csquotes}
\usepackage{quoting} 
\usepackage[color=orange]{todonotes}

\usepackage{tabularx}

\usepackage{multirow}

\usepackage{makecell}

\usepackage{array}

\usepackage{float}

\newcommand{\companyName}{Siemens AG}

\usepackage{tcolorbox}
\definecolor{amber}{rgb}{1.0, 0.75, 0.0}
\newtcolorbox{mybox}{colback=amber!10,colframe=amber}

\usepackage{siunitx}

\articletype{Research Article}

\begin{document}

\title{Adopting Microservices and DevOps in the Cyber-Physical Systems Domain: A Rapid Review and Case Study}

\author[1]{Jonas Fritzsch*}

\author[1]{Justus Bogner}

\author[1]{Markus Haug}

\author[3]{Ana Cristina Franco da Silva}

\author[2]{Carolin Rubner}

\author[2]{Matthias Saft}

\author[2]{Horst Sauer}

\author[1]{Stefan Wagner}

\authormark{Fritzsch \textsc{et al}}


\address[1]{\orgdiv{University of Stuttgart}, \orgname{Institute of Software Engineering}, \orgaddress{\state{Stuttgart}, \country{Germany}}}

\address[2]{\orgdiv{Siemens AG}, \orgname{Siemens Technology}, \orgaddress{\country{Germany}}}

\address[3]{Independent researcher, \country{Germany} (research participation while at $^1$)}

\corres{
    *Jonas Fritzsch\\
    \email{jonas.fritzsch@iste.uni-stuttgart.de}
}

\presentaddress{
    University of Stuttgart\\
    Institute of Software Engineering\\
    Universitätsstraße 38\\
    70569 Stuttgart\\
    Germany
}

\abstract[Summary]{
The domain of cyber-physical systems (CPS) has recently seen strong growth, e.g., due to the rise of the Internet of Things (IoT) in industrial domains, commonly referred to as \enquote{Industry 4.0}.
However, CPS challenges like the strong hardware focus can impact modern software development practices, especially in the context of modernizing legacy systems.
While microservices and DevOps have been widely studied for enterprise applications, there is insufficient coverage for the CPS domain.
Our goal is therefore to analyze the peculiarities of such systems regarding challenges and practices for using and migrating towards microservices and DevOps.
We conducted a rapid review based on 146 scientific papers, and subsequently validated our findings in an interview-based case study with 9 CPS professionals in different business units at \companyName.
The combined results picture the specifics of microservices and DevOps in the CPS domain.
While several differences were revealed that may require adapted methods, many challenges and practices are shared with typical enterprise applications.
Our study supports CPS researchers and practitioners with a summary of challenges, practices to address them, and research opportunities.
}



\maketitle

\section{Introduction}
The ongoing digital transformation of the industrial sector with a focus on Internet of Things (IoT) technologies is also referred to as \enquote{Industry 4.0}.\cite{Lasi2014}
In this context, industrial software engineering, i.e., the systematic development of high-quality cyber-physical systems (CPS), becomes more important than ever.\cite{Broy2014}
It is characterized by the digital representation and management of all involved industrial processes, with the goal to increase flexibility and evolvability.\cite{Ibarra-Junquera2021}
From a software engineering (SE) perspective, microservices and DevOps have similar objectives regarding these qualities.\cite{Waseem2020}
Even though both originated in the context of enterprise applications and are seen as the de facto standard for modern cloud-based software systems, they are also regarded as promising approaches for the IoT and CPS domain.\cite{Banica2018}
Within agile development, combining microservices and DevOps promises higher flexibility, shorter release cycles, and better maintainability, which are highly attractive benefits for CPS manufacturers.

Fundamental changes to the software architecture are usually accompanied by operational and organizational changes.
Hence, transforming a legacy system that is developed using a traditional process towards DevOps and microservices is a challenging task,\cite{Knoche2019} which needs to be carefully considered.\cite{Baskarada2018}
Various empirical studies cover challenges and practices for microservices and DevOps in the context of enterprise applications.\cite{Fritzsch2019d,Taibi2017a}
However, cyber-physical systems and the IoT domain differ from common enterprise systems in several aspects: 

\begin{itemize}
    \item Strong focus on hardware with a virtual representation (digital twin)~\cite{Broy2014}
    \item Distributed systems with heterogeneous communication mechanisms~\cite{Torngren2018a}
    \item Large amounts of generated data with real-time analysis requirements~\cite{Jiang2018}
    \item Strict regulations and requirements on safety and compliance~\cite{Li2018a}
\end{itemize}

\noindent
Existing microservices practices for enterprise applications might therefore be inappropriate for the CPS domain.
In this study, we aim to analyze differences in using microservices and DevOps in the CPS domain, especially when migrating legacy systems. The following research questions scope this objective:

\begin{enumerate}[label=RQ\arabic*:,leftmargin=1.45cm]
    \item What are \textbf{challenges} related to microservices and DevOps in the context of cyber-physical systems?
    \item What are \textbf{practices} related to microservices and DevOps in the context of cyber-physical systems?
    \item What are \textbf{rationales} why CPS professionals want to migrate their systems towards microservices and DevOps?
\end{enumerate}

\noindent
To this end, we formed an industry-academia collaboration and conducted a mixed-method empirical study.
First, we systematically analyzed scientific literature.
Second, we validated and extended the results through an interview-based case study with industry experts from different units at \companyName\footnote{\url{https://www.siemens.com}}, a global technology enterprise with extensive CPS expertise in several domains, ranging from manufacturing, healthcare, or mobility to energy.

\section{Background}
In this section, we briefly describe cyber-physical systems, why they evolve towards microservices and DevOps, as well as existing related work.

\subsection{Evolution of CPS towards Microservices and DevOps}
A major technological foundation for the ongoing Industry 4.0 transformation in the CPS domain is the Internet of Things (IoT),\cite{Mizutani2021} or Industrial Internet of Things (IIoT). 
Such systems are characterized by the integration of mechanical and electrical devices that represent smart endpoints.
Recent developments also utilize digital representations (digital twins), in particular for simulations in manufacturing systems.\cite{Ciavotta2020a} 
Gartner estimates the growth of IoT endpoints from 331.5 million in 2018 to 1.9 billion in 2028.\cite{Gartner2020}
This massive growth of involved devices poses new challenges for their efficient management.
Mass-customization and unpredictable workloads require re-configurable and highly automated platforms,\cite{Ciavotta2020a} which in turn has consequences for software architectures and development processes.
Combemale and Wimmer express the need for more comprehensive views on the development and operation of such environments.\cite{Combemale2020}

Microservices are fine-grained units forming a single application that are individually scalable and deployable, and offer more flexibility than a monolith.\cite{Fowler2015}
Enabled by technologies like containerization, cloud-native applications heavily rely on this architectural style.
DevOps comprises a variety of practices to connect development and operations in a seamless process.
These practices promote a high degree of agility and automation, which is key for efficiently operating microservice-based systems.
While it promises advantages for maintainability or scalability, the migration of existing systems towards microservices is a challenging task.\cite{Fritzsch2019d}
Existing research yielded many microservices refactoring approaches to mitigate effort and partly automate tasks.\cite{Fritzsch2019a,Ponce2019}
However, their applicability is heavily dependent on several factors like, e.g., the available input artifacts, targeted quality attributes, maturity of tool support, and its preceding evaluation.
Moreover, organizational challenges like restructuring teams and establishing a DevOps culture need to be considered as well and sometimes even pose a bigger challenge, especially for large organizations.\cite{Fritzsch2019d}
There are several studies available that collect empirical evidence on microservice migration in industry.
These studies mainly cover enterprise systems like retail or consumer services.
The CPS domain, however, has received little explicit attention yet for research on adopting microservices and DevOps.

\subsection{Related Work}
In their general CPS research roadmap, Törngren and Grogan discuss the inherent complexities of these systems, including causes, effects, and limitations of existing methodologies.\cite{Torngren2018}
Similarly, Fahmideh et al. acknowledge the lack of research on the development lifecycle for IoT systems and conducted a mixed-method study to derive a process framework.\cite{Fahmideh2021}
It covers essential tasks from a SE perspective, but does not focus on DevOps and microservices.

Campeanu contributed an early mapping study on microservices architectures in the context of IoT and cloud computing.\cite{Campeanu2018}
He cataloged 364 primary studies from 2015 to 2018, visualizing the rapid emergence of this technological combination.
In a more holistic review, Joseph et al. discuss the use of microservices in specific domains.\cite{Joseph2019}
They found 12 studies proposing specific solutions for the IoT domain, including related fields like healthcare or smart living.
The authors emphasize that \enquote{the major challenge faced in developing IoT solutions is having to deal with a plethora of devices that are heterogeneous in nature}.\cite{Joseph2019}
Similar findings were revealed in the review of 26 primary studies by Pereira et al. on the use of DevOps in IoT systems.\cite{Pereira2021a}
Heterogeneous device management was also found to be a main challenge.

Hasselbring et al. aim to mitigate lacking expertise in the adoption of Industry 4.0 solutions by proposing an approach to \enquote{introduce methods and culture of DevOps into industrial production environments},\cite{Hasselbring2019}  especially among small and medium-sized enterprises.
Taibi et al. focus on the migration aspect towards microservices and DevOps in general in their systematic mapping.\cite{Taibi2019b}
They identified common architectural principles and patterns for microservices from 23 analyzed case studies.
However, only one case was from the IoT domain.

In summary, several studies exist on either microservices and DevOps \textit{or} the CPS domain.
However, a comprehensive analysis of challenges and practices at the intersection of these three concepts is still missing.
We aim to close this gap, and provide foundations for specific approaches for modernizing existing CPS.

\section{Research Design}
\begin{figure}[h!]
    \centering
    \includegraphics[width=0.98\textwidth]{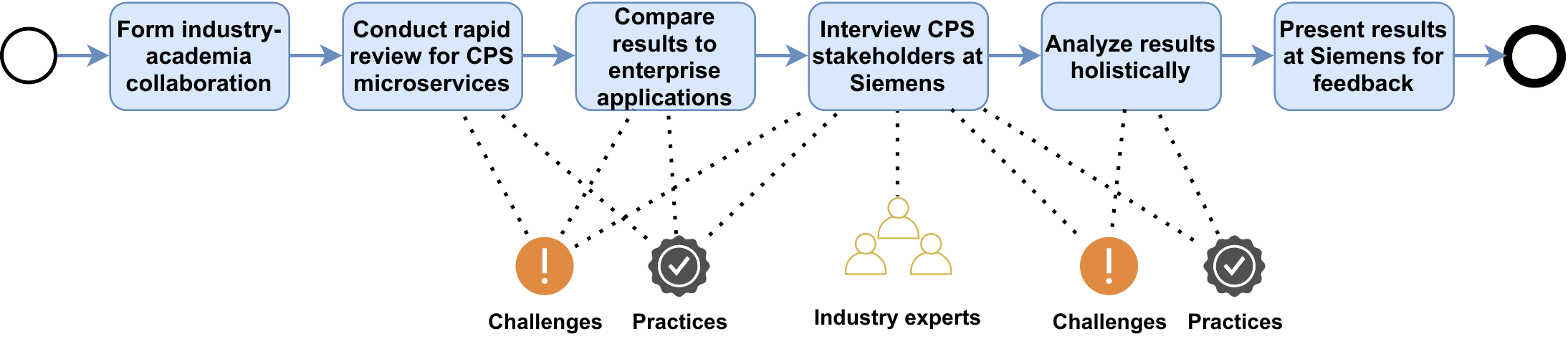}
    \caption{General research process}
    \label{fig:researchProcess}
\end{figure}

To answer the RQs, we formed a team of academic SE researchers and industry researchers from \companyName, who are experts on the CPS domain.
We decided to first conduct a literature review to analyze the scientific state of the art for microservices and DevOps in the context of CPS (see Section~\ref{meth:review}).
The results were then compared to similar publications on microservices in the enterprise application domain.
Afterwards, we conducted an interview-based case study at \companyName\ to compare the literature findings with industry experiences (see Section~\ref{meth:interviews}).
The results from both studies were holistically analyzed, discussed within the research team, and presented to a larger audience at \companyName\ for additional feedback.
This research process is visualized in Fig.~\ref{fig:researchProcess}.
We share the most important study artifacts as a digital appendix via Zenodo\footnote{\url{https://doi.org/10.5281/zenodo.6536538}}.

\subsection{Literature Review}
\label{meth:review}
The goal of a literature review is to identify and analyze existing publications for a certain topic, and to synthesize findings from the publication content.
Structured approaches to literature surveys like systematic literature reviews~\cite{Kitchenham2007} or systematic mapping studies~\cite{Petersen2008} provide a rigorous and reproducible process, but also require considerable effort.
Additionally, their results can be hard to integrate into industry practice.
We therefore decided to conduct a \textit{Rapid Review},\cite{Cartaxo2020} a lightweight secondary study performed in close collaboration with industry and focused on providing timely and easily consumable results.
The protocol of a rapid review is still systematic, but may consciously sacrifice rigor and extensiveness for industry relevance and efficiency.
The first four authors conducted the review, with the rest engaging in discussion.

\smallskip
\noindent
\textbf{Search Strategy}
By analyzing known relevant literature, discussing important keywords, and piloting a few alternatives, we finally decided to use the following two search strings, one for \enquote{microservices} and one for \enquote{DevOps}:

\smallskip
\begin{mybox}
    \begin{enumerate}
    \item \texttt{microservices $\land$ (cps $\lor$ cyber-physical $\lor$ manufacturing $\lor$ industrial $\lor$ edge)}
    
    \smallskip
    
    \item \texttt{devops $\land$ (cps $\lor$ cyber-physical $\lor$ manufacturing $\lor$ industrial $\lor$ edge)}
    \end{enumerate}
\end{mybox}
\smallskip

\noindent
An \texttt{AND} relation was used to combine each main term with CPS keywords, of which one had to match as well.
Both search strings were executed independently.
To reduce false positives, the query was only executed for the title field.
We selected the following five common digital libraries / academic search engines:

\begin{itemize}[noitemsep]
    \item Google Scholar
    \item ACM Digital Library
    \item IEEE Xplore
    \item ScienceDirect
    \item SpringerLink
\end{itemize}

\noindent
To ensure a manageable number of results, we defined a stopping criterion: only the first 50 publications per search string and source would be considered, i.e., a maximum of 500 publications (2 search strings $\times$ 5 sources $\times$ 50 hits).
After data extraction, one round of backward and forward snowballing~\cite{Wohlin2014} (reference and citation search) complemented this initial search.
The rationale for this strategy was to combine sufficiently broad coverage with manageable effort.

\smallskip
\noindent
\textbf{Study Selection}
We used the web-based tool Rayyan,\footnote{\url{https://rayyan.ai}} which offers convenient features to include or exclude studies in a multi-user mode.
Each paper was assigned to two researchers, who performed the selection independently and compared the results.
Disagreements were discussed within the whole group until consensus was reached.
Papers were \textbf{included} based on at least one of the following criteria:

\begin{itemize}[noitemsep]
    \item Paper describes CPS challenges associated with microservices or DevOps
    \item Paper describes a practice to develop CPS using microservices or DevOps
    \item Paper describes a practice to migrate CPS towards microservices or DevOps
\end{itemize}

\noindent
Practices include, e.g., processes, techniques (e.g., for system migration), platforms, reference architectures, algorithms, quality assurance practices, etc.
Additionally, a paper could also be \textbf{excluded} for one of the following reasons:

\begin{itemize}[noitemsep]
    \item Short paper with four pages or fewer
    \item Full-text not available to us
    \item Not published in English
    \item Published before 2015
    \item Not peer-reviewed scientific literature (e.g., books, student theses, etc.)
    \item Duplicate or extension (the more recent paper is kept)
\end{itemize}

\smallskip
\noindent
\textbf{Data Extraction}
Included papers were exported into a spreadsheet with columns for challenges and practices plus some additional columns for categorization, e.g., what part of the CPS domain the paper focused on.
To validate the sheet, all four researchers independently extracted the first five papers.
Afterwards, results were compared, which further improved the sheet and led to more concrete extraction guidelines.
We repeated this process a second time, after which we felt confident that our shared understanding leads to reasonably consistent extractions.
For efficiency, we therefore assigned each remaining paper to a single researcher for extraction.
If it was discovered during extraction that a paper did not completely fulfill the selection criteria, it was excluded.

\smallskip
\noindent
\textbf{Snowballing}
All successfully extracted papers were used in one round of snowballing.
For backwards snowballing, the references of all papers were automatically extracted from the PDF via a tool.\footnote{\url{https://github.com/helenocampos/PDFReferencesExtractor}}
The references were de-duplicated, and equally distributed among the four researchers for selection.
For forward snowballing, the included papers were again split up, and each researcher examined their citations via Google Scholar up to a maximum of 50 citations per paper.
All newly proposed papers were examined by the other researchers and only kept if nobody objected.
Afterwards, we extracted the snowballed papers following the same process as described above.

\smallskip
\noindent
\textbf{Data Synthesis}
The final extractions were then holistically analyzed to synthesize answers to our RQs.
For basic columns like the CPS domain or targeted SE activities, not much effort was required to consistently aggregate them.
However, since challenges and practices were extracted via direct paper quotes, we had to apply thematic synthesis.\cite{Cruzes2011}
We first assigned codes to each individual extraction (\textit{open coding}), and afterwards synthesized these codes into higher-level themes (\textit{axial coding}).
This was a very iterative process, with lots of discussion and the frequent merging or splitting of codes and themes.
The final codes and themes were aggregated and visualized using charts.

\subsection{Interview-Based Case Study}
\label{meth:interviews}
To validate and enrich the findings from the rapid review, we conducted an interview-based case study at \companyName.
A case study~\cite{Runeson2009} is a research method to deeply analyze something in a real-world setting (a \textit{case}), usually following a qualitative and exploratory methodology.
While a case study can provide rich insights into a topic, the generalizability of the produced results may be limited.
We conducted an \textit{embedded case study},\cite{Runeson2009} i.e., our \textit{case} was Siemens AG as a CPS manufacturer and our \textit{units of analysis} were individual cyber-physical systems in different organizations at Siemens.
These systems were studied through the experiences and opinions of their stakeholders, i.e., 
interview participants in various roles with sufficient industry experience (at least 3 years).
Ideally, their systems should also be in the process of migrating towards microservices and DevOps, or such a migration was considered for the future.
We used convenience sampling via our personal networks at \companyName\ to find matching participants.
For flexibility, we chose \textit{semi-structured interviews}.\cite{Seaman2008}
The interviews were conducted by the first three authors.

\smallskip
\noindent
\textbf{Study Preparation}
We created several artifacts for the interviews.
An \textit{interview preamble}~\cite{Runeson2009} explaining the process was used to recruit participants and to make them familiar with the study.
The preamble also described ethical considerations and requested consent for audio recordings.
Additionally, we created an \textit{interview guide}~\cite{Seaman2008} with the most important questions.
This guide was used to loosely structure the interviews.
We also prepared a small slide set with additional information for certain questions.
Lastly, we created a \textit{case characterization matrix}~\cite{Seaman2008} with basic attributes for the analysis.

\smallskip
\noindent
\textbf{Evidence Collection}
We conducted nine individual interviews, all of them via remote communication software.
Three interviews were held in English and six in German.
Interviews took 45 to 70 min, and all participants agreed to a recording.
At least two researchers were present during each interview, and swapped between moderator and note writer.
After a brief introduction, we showed a slide with microservices characteristics to ensure the same basic understanding.
We then inquired basic demographics, and let participants describe their system from the CPS domain.
Afterwards, we asked about challenges and practices related to microservices and DevOps, specifically in a migration context.
Additionally, we asked participants about their rationales for migrating their CPS towards microservices and DevOps.
To support participant reflections, we briefly introduced a migration methodology created by us,\footnote{The details of the methodology can be found in another paper from us.\cite{Fritzsch2022}} where participants could describe their experience in each step.

We manually transcribed each recording into a text document.
These documents were then sent to interviewees for review and approval.
Participants could delete sensitive content or adapt unclear statements.
We then used the approved transcripts for qualitative analysis.

\smallskip
\noindent
\textbf{Data Analysis}
To synthesize the interview findings, we used card-sorting,\cite{Zimmermann2016} a lightweight, collaborative analysis technique.
Instead of paper cards, we used a web-based tool with feedback cards for agile retrospectives.\footnote{\url{https://metroretro.io}}
Each interview block was assigned to one researcher, who then extracted relevant quotes per transcript into individual cards.
In an iterative process, these cards were then sorted according to their similarity, and preliminary labels were assigned.
Lastly, the results were presented to the other researchers.
Together, we refined the labels and formed higher-level categories.
Finally, important findings were extracted into spreadsheets, aggregated, and visualized.

\subsection{Data Availability}
For transparency and reproducibility, we share all important study artifacts for both the rapid review and the interviews on Zenodo.\footnote{\url{https://doi.org/10.5281/zenodo.6536538}}
This includes, e.g., the list of primary studies (CSV), the extracted practices and challenges (CSV), the interview guide, and the aggregated interview results (CSV).

\section{Results}
In this section, we first present the results for the rapid review (challenges and practices) and then for the interviews (rationales, challenges, and practices).

\subsection{Rapid Review}
\begin{figure}[h!]
    \centering
    \includegraphics[width=0.75\textwidth]{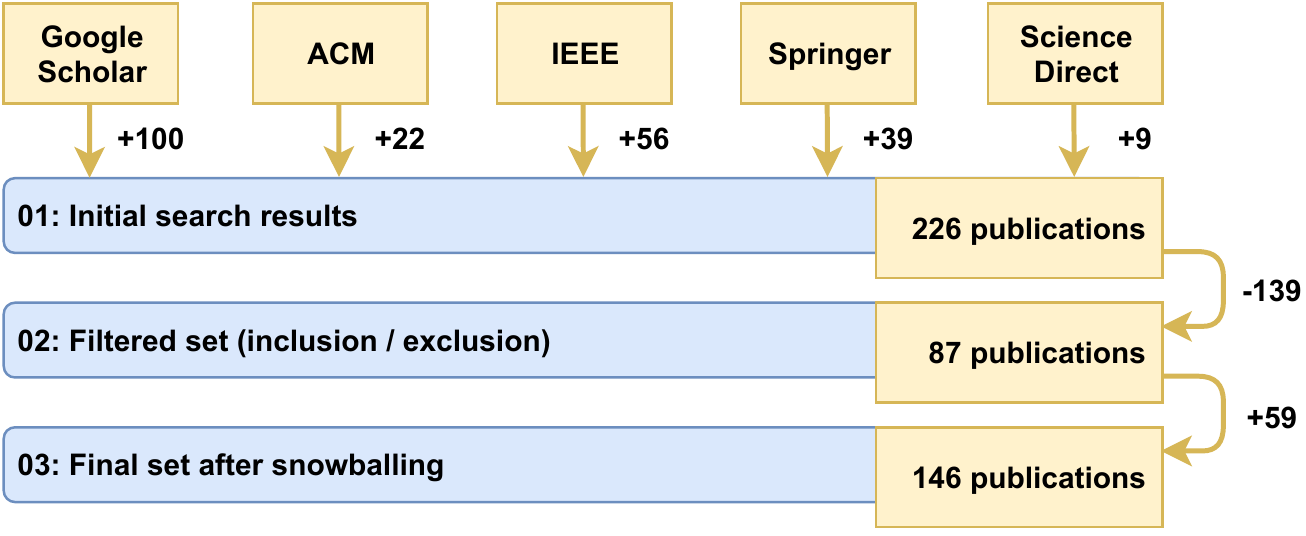}
    \caption{Literature review: number of publications per stage}
    \label{fig:reviewResultStages}
\end{figure}

In total, we included 146 primary studies in the review, 87 of which were discovered in the initial search and another 59 during snowballing (see Fig.~\ref{fig:reviewResultStages}).
The papers were published between 2015 and 2021, and covered all phases of the software engineering lifecycle, with \emph{design} being the major focus (mentions in 127 papers), followed by \textit{deployment} (44) and \textit{operation} (31).

\begin{figure}[h]
    \centering
    \includegraphics[width=0.96\textwidth]{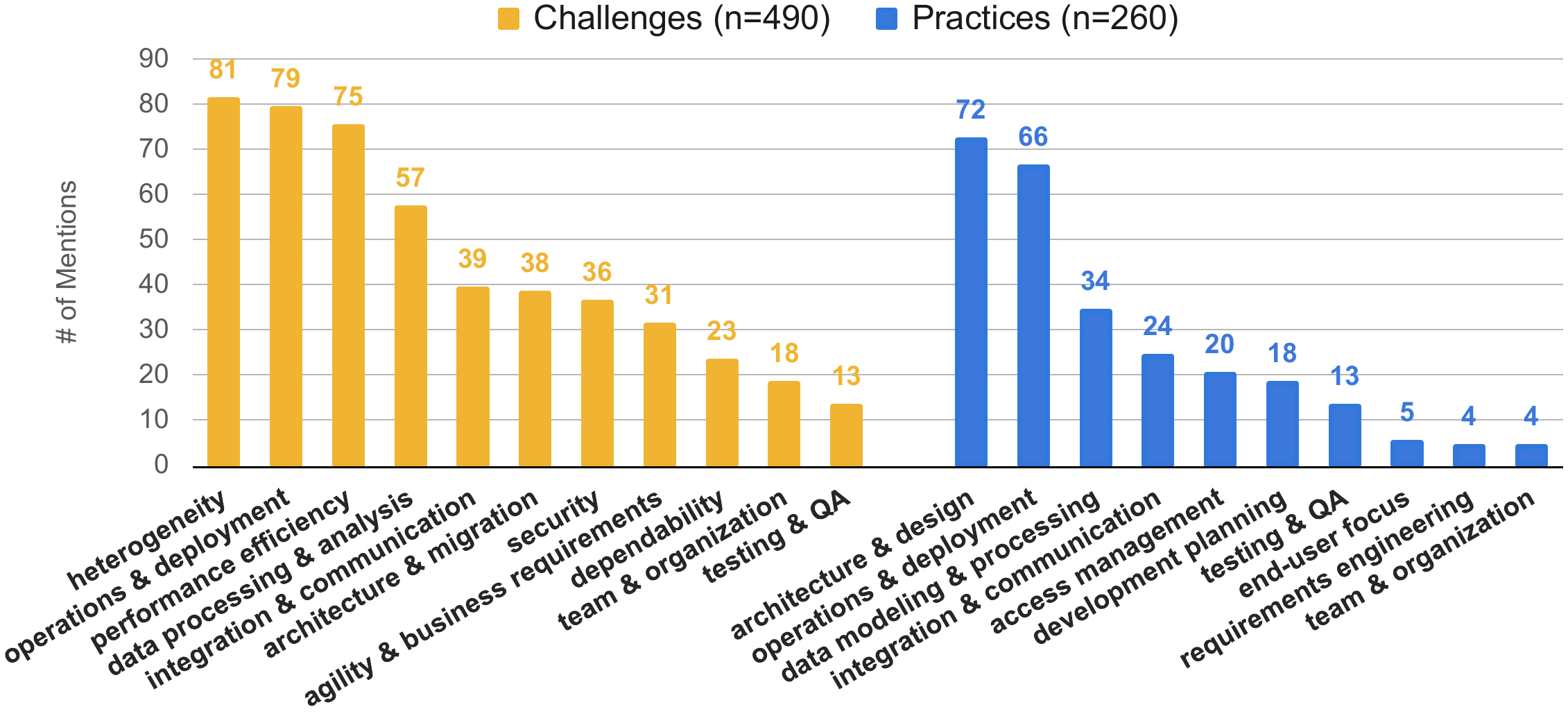}
    \caption{Literature review: challenge and practice categories}
    \label{fig:reviewCategories}
\end{figure}

\subsubsection{Challenges (RQ1)}
From all 146 primary studies, we extracted 405 statements, which were converted into 46 challenge labels in 11 categories during the analysis.
The 46 challenges were mentioned a total of 490 times among all statements.
A challenge was counted at most once per paper.
Fig.~\ref{fig:reviewCategories} shows the 11 challenge categories with their number of mentions as orange bars, while Table~\ref{tab:rr-challengens-practices} shows the 10 most frequent challenge labels. 

The dominant category was \textit{heterogeneity}, with \textit{technological heterogeneity of devices} (48) and \textit{variety of data formats and communication protocols} (24) being the most mentioned challenges in it.
Numerous devices also pose challenges for \textit{operations \& deployment}.
In this category, the management of components like the \textit{efficient placement, scheduling, and load balancing of microservices} (18) were prevalent issues. 
The third most popular category was \textit{performance efficiency}, which contained the second and third most mentioned challenges overall: \textit{realtime processing requirements} (31) and \textit{limited hardware resources of devices and edge nodes} (29).

The \textit{large amount of data to process} (25) together with \textit{complex data storing and processing requirements} (18) highlight prevalent challenges for \textit{data management \& processing}.
In general, we observed a strong tendency towards hardware-related aspects and the management and operation of a multitude of distributed components. 
One of the major aspects here was \textit{establishing connectivity and interoperability} (19), which was the most mentioned challenge in the \textit{integration \& communication} category.
The \textit{architecture \& design} category was mostly driven by the \textit{limited evolvability of traditional architectures like monoliths and SOA} (21), which expresses the need for more suitable architectural patterns.

The \textit{security} category with issues around \textit{data privacy and authorization for a multitude of components} (12) and \textit{strong requirements for security} (9) also received considerable attention.
\textit{Agility \& business requirements} pose challenges on CPS as well, as the two most frequent challenges in this category reflect: \textit{highly dynamic and uncertain environments} (10) and \textit{quickly emerging technological advancements require more flexibility} (8).
The \textit{dependability} category was dominated by challenges around \textit{strict safety and compliance regulations} (10) and \textit{strong requirements for reliability} (8).
Organizational and process-related challenges, however, played a minor role. The most mentioned challenge in this category was \textit{adapting IoT/CPS processes and philosophy to DevOps} (6).

\subsubsection{Practices (RQ2)}
For practices, we extracted 283 statements, which were condensed into 32 labels in 10 categories. 
The 32 practices were mentioned a total of 260 times among all statements.
 
As with challenges, a practice was counted at most once per paper.
The first of two dominant categories was \textit{architecture \& design}, with specific \textit{architecture proposals for various application contexts} (55) being the most frequent label, e.g., microservice architectures for industrial control systems (ICS) or cloud robotics.
Other practices in this category were rather insignificant.
The second dominant practice category was \textit{operations \& deployment}, with the top label being \textit{techniques for placement, scheduling, allocation, deployment in various domains} (35), often for edge-cloud continuums.
Other labels received considerably fewer mentions: the second most addressed practice was targeting \emph{techniques or patterns for monitoring of microservice-based systems} (12).

All other categories received notably less attention.
\emph{Data modelling \& processing} was the third most popular category, with practices centered around \emph{microservices for data processing and analytics on IoT/edge nodes} (18) or \emph{conceptual modeling \& model-driven engineering} (16).
Several papers also proposed practices for \emph{integration \& communication} aspects, such as \emph{techniques and patterns for microservice registration and discovery in IoT/IIoT environments} (7), followed by techniques for \emph{autonomous / automated device management and reconfiguration} (5).
\emph{Access management} centered around various \textit{security techniques for IoT/edge systems} (16).
\emph{Development planning} was dominated by practices for \emph{DevOps adoption in various application contexts} (15), such as ICS or IoT.
While several DevOps-related aspects are also covered by other categories and labels, this label focuses on more general, process-related facets.
Finally, \emph{testing \& QA} solutions received relatively little attention (13), similar as with challenges in this area.
Most practices in this category were targeting simulation and performance testing.

In general, practices were more condensed within a few categories, dominated by concrete architecture proposals and techniques for the dynamic scheduling of microservices within edge/fog/cloud computing.
Organizational and process-related solutions received noticeably little attention in our review. 

\begin{table*}[h]
	\setlength{\tabcolsep}{4pt}
    \centering
    \caption{Most mentioned challenges and practices (by number of papers)}
    \label{tab:rr-challengens-practices}
    \begin{tabular}{
	    >{\raggedright\arraybackslash}p{0.7\textwidth}
	    >{\raggedleft\arraybackslash}p{0.02\textwidth}
	    >{\raggedright\arraybackslash}p{0.235\textwidth}
	    }
        \toprule
        \textbf{Challenge} & \textbf{\#} & \textbf{Category} \\
        \midrule
        technological heterogeneity of devices & 48 & heterogeneity \\
        realtime processing requirements (response time of cloud microservices too high) & 31 & performance efficiency \\
        limited hardware resources of devices and edge nodes & 29 & performance efficiency \\
        large amount of data to process & 25 & data processing \& analysis \\
        variety of data formats and communication protocols & 24 & heterogeneity \\
        evolvability is limited by traditional architectures (monoliths, SOA) & 21 & architecture \& design \\
        efficient management of a large number of components (microservices, devices, etc.) & 20 & operations \& deployment \\
        establishing connectivity and interoperability & 19 & integration \& communication \\
        complex data storing and processing requirements & 18 & data processing \& analysis \\
        efficient microservices placement, scheduling, and load balancing & 18 & operations \& deployment \\
        \midrule
        \textbf{Practice} \\
        \midrule
        architecture proposals for various application contexts (ICS, IIoT, cloud robotics, healthcare, manufacturing, etc.) & 55 & architecture \& design \\
        techniques for placement, scheduling, allocation, deployment in various domains & 35 & operations \& deployment \\
        microservices for data processing and analytics on IoT/edge nodes & 18 & data modelling \& processing \\
        conceptual modeling and model-driven engineering & 16 & data modelling \& processing \\
        security techniques for IoT/edge systems & 16 & access management \\
        DevOps adoption in various application contexts (CPS, ICS, IoT) & 15 & DevOps process \\
        techniques \& patterns for monitoring of microservice-based systems & 12 & operations \& deployment \\
        techniques, tools, patterns for efficient cloud/edge deployment & 8 & operations \& deployment \\
        techniques \& patterns for microservice registration and discovery & 7 & integration \& communication \\
        guidelines for determining microservices boundaries & 6 & architecture \& design \\
        microservices architectural style to meet requirements of IoT/edge gateways & 6 & architecture \& design \\
        orchestration of services / containers for automated deployment in IoT/edge solutions & 6 & operations \& deployment \\
        simulation of IoT environments / applications to facilitate testing and evaluation & 6 & testing \& QA \\
        \bottomrule
    \end{tabular}
\end{table*}

\subsection{Interviews}
In total, we conducted nine interviews with participants from different business units across \companyName.
Table~\ref{tab:participants} lists their demographics.
The experience column gives their professional experience in years, with the amount of microservice experience in parentheses.
Except for P7, all participants had more than 10 years of professional experience.
However, their prior experience with microservices and DevOps varies considerably, with two participants with no prior experience, and two to six years for the rest.
The interviewees described seven different systems across a variety of fields.
Table~\ref{tab:systems} shows their purpose, size, staffing, and the projected migration timeframe.
As we can see, even the \enquote{smallest} system (S6) still had over \num{100000} lines of source code.

\begin{table*}[h]
    \centering
    \caption{Participant demographics (years of microservices experience in parentheses)}
    \label{tab:participants}
    \begin{tabular}{l l r c}
        \toprule
        \textbf{ID} & \textbf{Role}             & \textbf{Experience} & \textbf{System} \\
        \midrule
        P1          & architect, researcher     &              18 (3) & S1              \\
        P2          & architect                 &              19 (3) & S2              \\
        P3          & software development lead &              20 (6) & S3              \\
        P4          & architect                 &              15 (2) & S4              \\
        P5          & architect                 &              25 (0) & S5              \\
        P6          & architect                 &              16 (0) & S4              \\
        P7          & researcher                &              6  (5) & S6              \\
        P8          & project manager           &              14 (4) & S6              \\
        P9          & project manager           &              15 (5) & S7              \\
        \bottomrule
    \end{tabular}
\end{table*}

\begin{table*}[h]
	\setlength{\tabcolsep}{4pt}
	\centering
    \caption{System and migration properties}%
	\label{tab:systems}
    \begin{tabular}{
	    >{\raggedright\arraybackslash}p{0.04\textwidth}
	    >{\raggedright\arraybackslash}p{0.32\textwidth}
	    >{\raggedright\arraybackslash}p{0.09\textwidth}
	    >{\raggedright\arraybackslash}p{0.12\textwidth}
        >{\raggedright\arraybackslash}p{0.1\textwidth}
        >{\raggedright\arraybackslash}p{0.24\textwidth}
	}
        \toprule
        \textbf{ID} & \textbf{Purpose}                                       & \textbf{Size} & \textbf{Staffing}       & \textbf{Time\-frame} & \textbf{Targeted Architecture} \\
        \midrule
        S1          & manufacturing management system               & n/a           & 50 (5~teams)   & 2020--               & aspects of microservices       \\
        S2          & engineering of automation systems                      & 70 MLOC       & 300                     & 2018--2026           & aspects of microservices       \\
        S3          & digital healthcare services platform                   & 100 MS        & 250 (25~teams) & 2017--2023           & full microservice architecture \\
        S4          & digital healthcare services platform                   & 15 MLOC       & 300 (30~teams) & 2019--               & full microservice architecture \\
        S5          & control software for electric motors          & 5 MLOC        & 100                     & 2011--               & modularization                 \\
        S6          & digital twin platform for power distribution equipment & 150 kLOC      & 10                      & 2021--               & full microservice architecture \\
        S7          & process control engineering                            & 600 kLOC      & 250 (12~teams) & 2021--2023           & aspects of microservices       \\
        \bottomrule
    \end{tabular}
\end{table*}

All systems were in the process of a migration.
However, the expected migration timeframes differed considerably, ranging from two years (S7) to over a decade (S5).
Often, no distinct end date was mentioned, e.g., for S1, S4, or S6.
For some systems, the reason was an ongoing evaluation of the feasibility of a full migration.
In other cases, the migration followed a continuous evolution strategy, where components were slowly migrated towards microservices when they had to be changed in the course of day-to-day development.
We also discussed the architecture that should be achieved with the migration, which varied regarding the commonly understood characteristics of microservices,\cite{Fowler2015} i.e., the characteristics we introduced at the beginning of the interviews.
For three systems, the targeted architecture was a full microservice architecture that relied on almost all characteristics.
For another three systems (denoted with \textit{aspects of microservices} in Table~\ref{tab:systems}), the targeted architecture fulfilled some microservices characteristics, but not enough to be considered a full microservice architecture.
That means S5 was the only system where the team mainly aimed to achieve a higher modularization.
All other systems set out for a partial (i.e., some characteristics) or full microservice architecture (i.e., nearly all characteristics).

\subsubsection{Migration Rationales (RQ3)}

\begin{figure}[h!]
    \advance\leftskip0.38cm
    \includegraphics[width=0.91\textwidth]{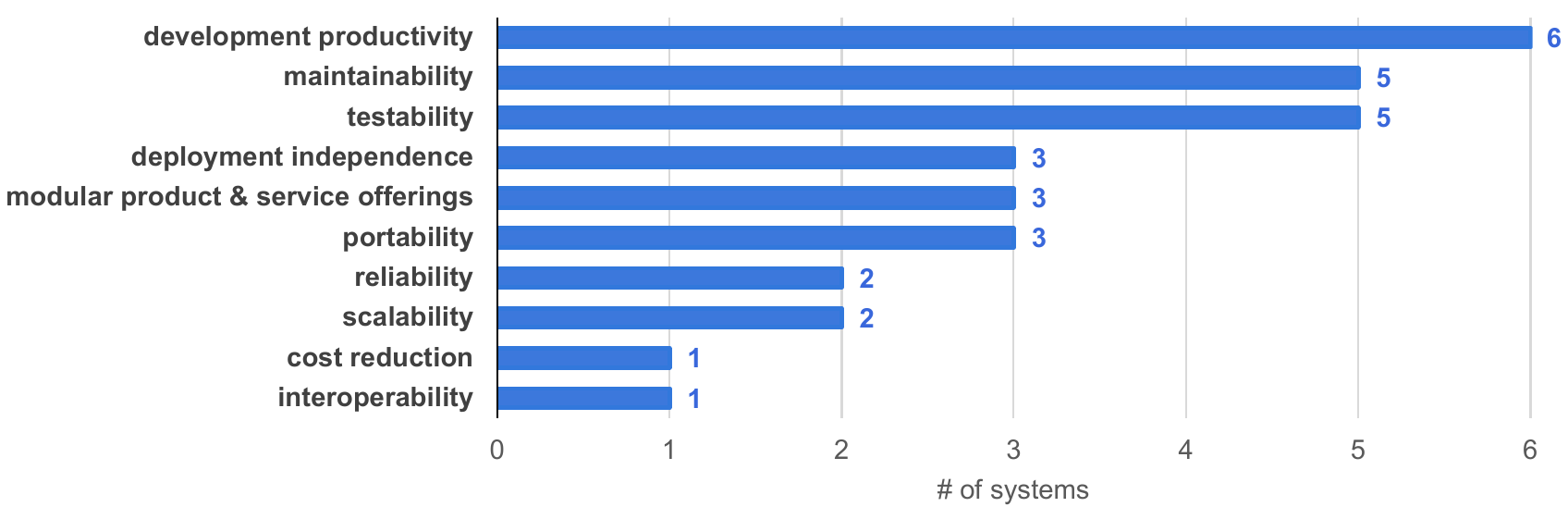}
    \caption{Distribution of interview migration rationales}
    \label{fig:intMigrationRationales}
\end{figure}

When asked for their migration rationales, i.e., why they modernized their systems towards microservices and DevOps, our participants mentioned a total of 10 different reasons, most of them quality attribute goals (see Fig.~\ref{fig:intMigrationRationales}).
For six of the seven systems, increased \textit{development productivity} was a main driver (all systems except S1).
Decreasing cycle time, establishing faster feedback loops, or providing new functionality faster were prevalent themes here.
Related to this was the goal to improve \textit{maintainability}, which was mentioned for five systems (S2, S4, S5, S6, S7).
While better maintainability ultimately should lead to better productivity, the mentioned reasons here were more technical and structural, e.g., improving modularity, reducing coupling, improving code quality, reducing complexity, or improving reusability.
Five systems (S2, S3, S4, S5, S7) also had the goal to improve \textit{testability}, i.e., to be able to test smaller components in isolation and in less time.

Other rationales occurred less frequently.
For three systems (S3, S4, S6), \textit{deployment independence} was explicitly mentioned, i.e., each team should be able to release and deploy their services independently of all others.
More a business than a technical driver was the need for \textit{modular product \& service offerings}.
This was named for S1, S2, and S4, where fine-grained and more specialized offerings to customers and avoiding redundant offerings were very important.
In three cases (S1, S2, S6), \textit{portability} was mentioned as a migration rationale, i.e., supporting various deployment environments like cloud vs. edge, Windows vs. Linux, digital twin platform vs. real hardware, etc.
Less prevalent rationales were improving \textit{reliability} (S4, S6) or \textit{scalability} (S4, S6), even though these two are typical quality attributes associated with microservices.
Lastly, \textit{cost reduction} (S6) and \textit{interoperability} (S2) were only mentioned once.

\subsubsection{Challenges (RQ1)}
\begin{figure}[h!]
    \centering
    \includegraphics[width=0.88\textwidth]{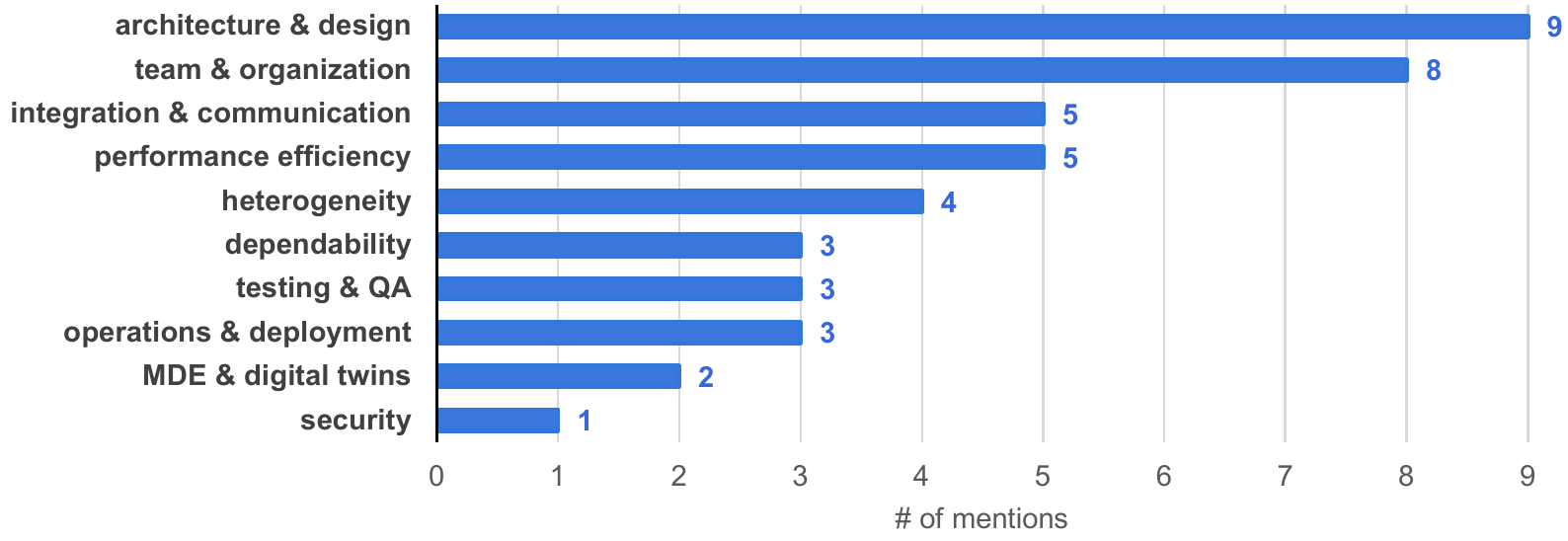}
    \caption{Distribution of interview challenge categories}
    \label{fig:intChallengeCategories}
\end{figure}

Our interview analysis revealed 26 unique challenge labels, from which we could reuse 16 from the literature review, i.e., 10 were newly created (38\%).
These individual challenges were grouped into 10 categories, all of them reused from the review.
The most prominent categories were \textit{architecture \& design} (9 mentions), \textit{team \& organization} (8), \textit{integration \& communication} (5), and \textit{performance efficiency} (5).
Categories like \textit{operations \& deployment} (3), \textit{MDE \& digital twins} (2), or \textit{security} (1) were less prominent.
Furthermore, individual labels were pretty spread out, with 13 challenges being mentioned only once and no challenge being mentioned for more than 3 of the 7 systems, i.e., there were no \enquote{top} challenges affecting most CPS in our sample.
The category distributions are shown in Fig.~\ref{fig:intChallengeCategories}, while the most mentioned challenges are presented in Table~\ref{tab:intTopChallenges}.
For the top categories, we present further details below.

\begin{table*}[h]
    \centering
    \caption{Challenges affecting 2 or 3 systems}%
    \label{tab:intTopChallenges}
    \begin{tabular}{
        lclc
	    }
        \toprule
        \textbf{Challenge} & \textbf{\# of Systems} & \textbf{Category} & \textbf{in LR} \\
        \midrule
        reducing coupling / hidden dependencies & 3 & architecture \& design & \textbf{no} \\
        establishing connectivity and interoperability & 3 & integration \& communication & yes \\
        realtime processing requirements & 3 & performance efficiency & yes \\
        lack of personnel with modern technology experience & 3 & team \& organization & yes \\
        technical debt of legacy system & 2 & architecture \& design & \textbf{no} \\
        strong requirements for reliability & 2 & dependability & yes \\
        technological heterogeneity of devices & 2 & heterogeneity & yes \\
        variety of data formats and communication protocols & 2 & heterogeneity & yes \\
        predictability and consistency in distributed processes & 2 & integration \& communication & yes \\
        limited hardware resources of devices and edge nodes & 2 & performance efficiency & yes \\
        building developer confidence in microservices \& DevOps & 2 & team \& organization & yes \\
        creating shared system understanding with all teams & 2 & team \& organization & \textbf{no} \\
        creating new decoupled test suites & 2 & testing \& QA & \textbf{no} \\
        \bottomrule
    \end{tabular}
\end{table*}

\paragraph{Architecture \& Design}
All systems except S1 reported at least one challenge from this category.
The most mentioned label was \textit{reducing coupling or hidden dependencies} (S3, S4, S6), i.e., participants found it challenging to untangle elements in preparation for the migration or service decomposition.
Another challenge in this area was \textit{technical debt of legacy systems} (S2, S5), which would slow down a migration considerably.
All other challenges like \textit{lack of architecture documentation} (S6), \textit{massive system size} (S2), or \textit{incorporating legacy components} (S6) were only mentioned once.
Lastly, only one label (\textit{limited evolvability due to traditional architectures like monoliths or SOA} from S7) was reused from the review, as most mentioned challenges here were not strongly specific to CPS.

\paragraph{Team \& Organization}
Challenges related to human factors were also very prominent.
For S4, S5, and S7, the \textit{lack of personnel with modern technology experience} was problematic, i.e., many CPS engineers would know little about microservices or DevOps.
Project manager P9 remarked that \enquote{architects with the required knowledge are currently most wanted in the project and are therefore somewhat of a bottleneck}.
Related to that, \textit{building developer confidence in microservices and DevOps} was sometimes challenging (S5, S6), and could lead to \enquote{trust issues} (researcher P7) if not properly managed.
For very large systems, a challenge not mentioned in the literature was \textit{creating a shared system understanding with all teams} (S2, S4).
Lastly, \textit{adapting the CPS processes and philosophy to microservices and DevOps} was problematic for S7, e.g., P9 saw the need for a cultural change while moving to DevOps and cross-functional teams in his department.

\paragraph{Integration \& Communication}
This category was only comprised of two labels.
However, they were mentioned for two and three systems respectively, and both appeared in the literature review.
For S1, S4, and S6, \textit{establishing connectivity and interoperability} between the newly distributed services was difficult.
Architect P1 saw \enquote{the challenging aspects of communication via message-passing} as the main problem in the migration.
Related to this was the challenge to \textit{maintain predictability and data consistency within distributed processes} (S1, S4).
Eventual consistency was not acceptable in some use cases, so creating \enquote{several services that are independent yet for which we also need to guarantee data consistency} (P6) was difficult and therefore sometimes completely avoided.

\paragraph{Performance Efficiency}
In this category, two challenges were mentioned, both of which could be reused from the literature review.
As expected, several CPS (S1, S4, S5) had strong \textit{real-time processing requirements}, which needed to be kept intact during and after the migration.
This meant that the response time of cloud-based microservices was often not sufficient, e.g., for \enquote{functionalities that need to be close to the production line} (P1).
For some use cases, interprocess communication between microservices was not feasible at all, or as described by architect P5: \enquote{we need to calculate cycles for electric current regulators within a two-digit microsecond window.}
A related challenge were the \textit{limited hardware resources of devices and edge nodes} (S4, S5).
This could also impact latency or scalability, but mainly limited using resource-intensive technologies like containerization.

\subsubsection{Practices (RQ2)}
\begin{figure}[h!]
    \advance\leftskip1.5cm
    \includegraphics[width=0.76\textwidth]{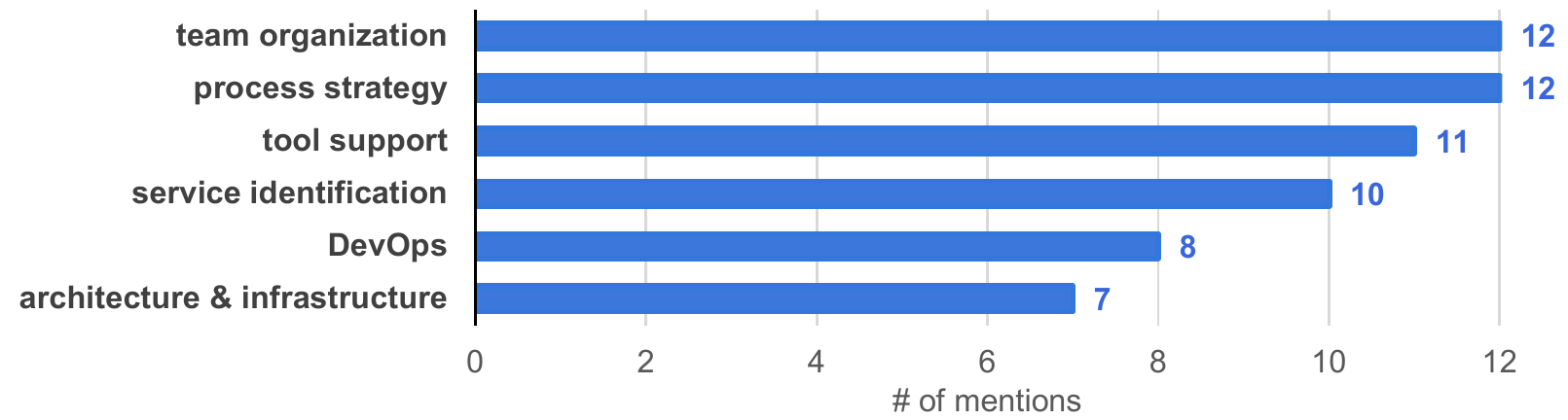}
    \caption{Distribution of interview practice categories}
    \label{fig:intPracticeCategories}
\end{figure}

We identified practices in the following categories, each listed with the number of mentioned practices: \textit{team organization} (12 mentions), \textit{process strategy} (12), \textit{tool support} (11), \textit{service identification} (10), \textit{DevOps} (8), and \textit{architecture \& infrastructure} (7). As we provided the interviewees with a stepwise migration methodology
created by us, the practice categories were prescribed to some extent. Fig.~\ref{fig:intPracticeCategories} shows the categories we queried about with the number of mentions respectively. We discuss them in detail below.

\paragraph{Team Organization \& DevOps} Six participants reported an associated team re-organization during their migration.
The prevalent motivation was promoting team independence and fostering collaboration between teams,  mentioned four times.
Two participants reported no organizational changes (P5, P6), and also stated that establishing DevOps practices was not a priority yet.
Reported DevOps practices were establishing CI/CD pipelines (S1, S2, S3, S6) and using cloud platform services (S1, S3).
Four interviewees had already established a DevOps culture.
Two aimed to do so (S4, S7), but were hindered by technical or organizational obstacles.
One participant remarked that team re-organization was a sensitive topic in Germany, and related it to the difficulty of DevOps adoption in large enterprises.

\paragraph{Process Strategy} The migration of a legacy system can follow different strategies.
For brownfield developments, we distinguish between a \textit{re-build} and \textit{re-factor} type.
The former is often used when the system is rebuilt using newer technologies and was applied for systems S1, S2, and S4.
Sometimes, even a mix of several strategies can be used for different parts or subsystems, as was the case for S2, S4, and S6.
We can further distinguish between a \textit{big bang} migration that aims to minimize the duration, and a continuous evolution approach that tries to minimize the needed resources.
The latter was used for four systems (S4, S5, S6, S7), and often based on the popular \textit{Strangler} pattern~\cite{Fowler2004} (S2, S3, S4).
Gradually building the new system with parallel operation next to the existing one was reported for S4 and S6.
As rationales, participants stated limited resources, time pressure, and strong requirements for stability and compliance.

\paragraph{Service Identification \& Tool Support} Existing literature identifies the decomposition task as the most complex one in a migration scenario.\cite{Fritzsch2019d}
In a similar notion, P1 stated that \enquote{success or failure will depend on the services having good granularity or not}.
Five of our participants described service decomposition as a manual task.
P1 and P3 reported Domain-Driven Design as the used technique, others mentioned static code and metadata analysis (P2, P9), as well as dynamic analysis (P5) of the existing system.
Our participants used a variety of mostly commercial tools that support the system comprehension via static code analysis (P2, P3, P6, P8).
However, a subsequent automatic decomposition is often beyond their scope.
For this purpose, P5, P6, and P9 relied on self-developed tools, two involved external consultants to compensate missing in-house expertise (P1, P5), and another two did not use any tools for this purpose (P4, P7).

\paragraph{Architecture \& Infrastructure} Three participants mentioned workshops as a practice to support the architecture design and infrastructure choices (P2, P3, P4).
Conducted cross-team and involving all system stakeholders, they facilitate the build-up of a common understanding and the identification of requirements.
P6 expressed his interpretation of agility in the following way: \enquote{Big upfront design is a no-go. Just as much architecture as necessary.}
For the preferred cloud model, platform as a service (PaaS) hosting was mentioned twice.
On-premise hosting was a strong requirement for S7 due to its tight integration with physical devices.

\section{Discussion}
The review and interview results show many similarities, but also reveal certain discrepancies.
In the following, we first compare the CPS results of our rapid review to similar literature studies or surveys for enterprise applications.
Afterwards, we contrast the rapid review results with the interview results.

\subsection{Comparison with Enterprise Application Microservices}

\begin{figure}[h!]
    \centering
    \includegraphics[width=0.85\textwidth]{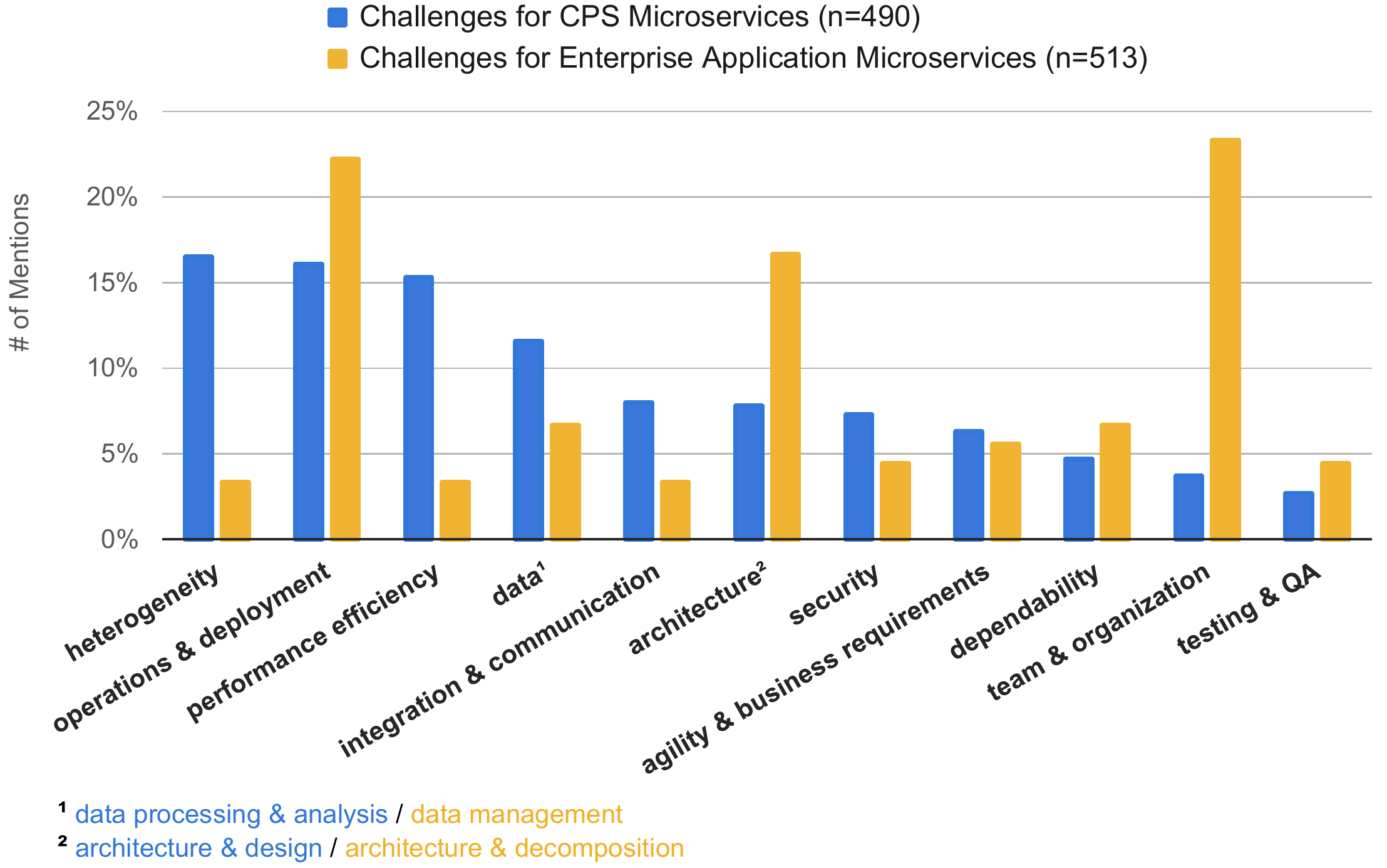}
    \caption{Differences in reported microservices challenges between CPS and enterprise applications domains}
    \label{fig:ChartChallenges}
\end{figure}

\begin{figure}[h!]
    \centering
    \includegraphics[width=0.78\textwidth]{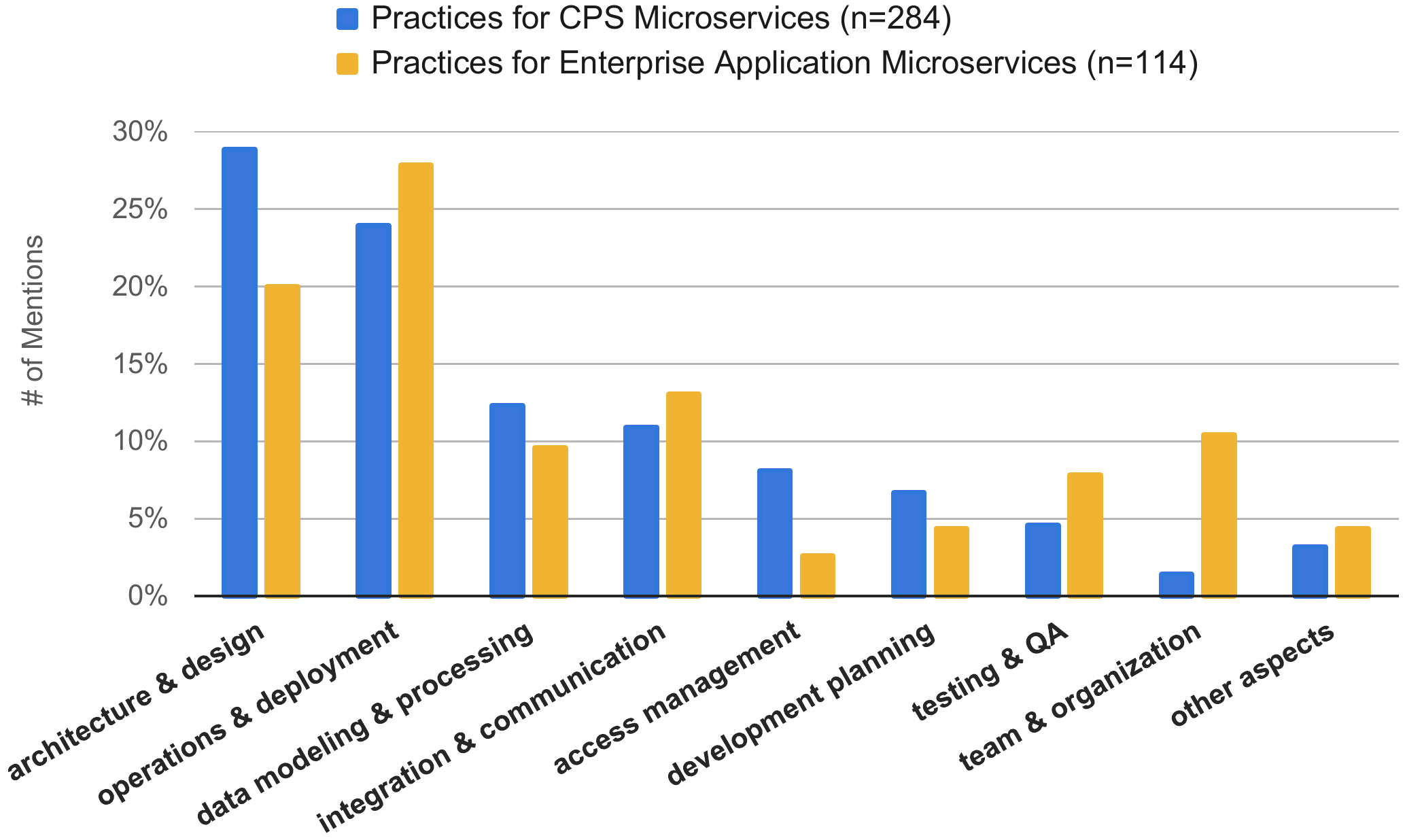}
    \caption{Differences in reported microservices practices  between CPS and enterprise applications domains}
    \label{fig:ChartPractices}
\end{figure}

To highlight the peculiarities of the CPS domain and show differences to common enterprise applications, we extracted challenges and practices from existing studies on enterprise application microservices.
For this purpose, we relied on our extensive experience with enterprise application microservices in industry~\cite{Fritzsch2019d,Bogner2019-ICSA,Matias2020,Bogner2021a,Vale2022} to select and analyze nine publications, primarily recent literature studies and two interview studies, that focus on general practices and challenges in the microservices and DevOps migration context.
Table \ref{table:ExistingReviews} lists these nine publications along with the number of included primary sources or study participants.
The last two columns show the number of extracted challenges and practices per paper.
Altogether, we extracted 513 challenges and 114 practices from all sources. 
For reviews, we counted a challenge or practice per primary source that stated it, for the interview and survey paper per participant that stated it.
To ease comparability, the labeling was harmonized with our preceding literature review by using the same categories, as can be seen in Figs.~\ref{fig:ChartChallenges} and \ref{fig:ChartPractices}.

\begin{table*}[ht]
    \centering
	\caption{Empirical studies and reviews on enterprise application microservices}
	\label{table:ExistingReviews}
	\begin{tabular}{
	    p{0.015\textwidth}
	    >{\raggedright\arraybackslash}p{0.51\textwidth}
	    >{\raggedleft\arraybackslash}p{0.03\textwidth}
	    >{\raggedleft\arraybackslash}p{0.08\textwidth}
	    >{\raggedleft\arraybackslash}p{0.08\textwidth}
	    >{\raggedleft\arraybackslash}p{0.06\textwidth}
	    >{\raggedleft\arraybackslash}p{0.06\textwidth}
	}
		\toprule
		\textbf{ID} & \textbf{Title} & \textbf{Year} & \textbf{Type} & \textbf{\# Sources} & \textbf{\# Chal.} & \textbf{\# Prac.}\\ 
        \midrule
		1 & Deployment and communication patterns in microservice architectures: A systematic literature review & 2021 & SLR & 38 & 57 & 14 \\
		2 & Monoliths to microservices - Migration Problems and Challenges: A SMS & 2021 & SMS & 37 & 63 & 2 \\
		3 & Understanding and addressing quality attributes of microservices architecture: A Systematic literature review & 2021 & SLR & 72 & 0 & 20 \\
		4 & A Systematic Mapping Study on Microservices Architecture in DevOps & 2020 & SMS & 47 & 72 & 37 \\
		5 & Promises and challenges of microservices: an exploratory study & 2020 & Interview + Survey & 21+37 & 111 & 21 \\
		6 & Architecting with microservices: A systematic mapping study & 2019 & SMS & 103 & 0 & 7 \\
		7 & Microservices Migration in Industry: Intentions, Strategies, and Challenges & 2019 & Interview & 16 & 70 & 0 \\
		8 & Continuous Integration, Delivery and Deployment: A Systematic Review on Approaches, Tools, Challenges and Practices & 2017 & SLR & 69 & 140 & 13 \\
		\midrule
		Total & & & & 440 & 513 & 114 \\
		\bottomrule
	\end{tabular}
\end{table*}

\paragraph{Challenges}
\textit{Heterogeneity} reflects the most mentioned challenge with CPS systems.
It goes along with achieving sufficient system \textit{performannce} as well as efficient \textit{data processing} in such environments, which are the third and forth most mentioned challenges for CPS.
We attribute these aspects to typical IoT/IIoT sytems, which are characterized by operating a large number of devices with limited hardware resources.
Compared to enterprise applications, these three challenge categories, plus \textit{integration \& communication} aspects, are seen as more important by the CPS literature.
Challenges with enterprise systems, on the other hand, are strongly centered around \textit{organizational aspects}.
This category seems to have received very little attention in the CPS literature, as the big gap in mentions reveals.
Difficulties with efficient \textit{operations \& deployment} follow, which is a very relevant category also for CPS systems.
When it comes to \textit{architectural aspects}, the gap again reveals a higher importance for typical enterprise systems.
Here, challenges are mainly centered around finding a suitable decomposition into microservices.
As many CPS are naturally designed as distributed systems, the modularization aspect might therefore be perceived as less challenging compared to other design aspects.
\textit{Security concerns} showed a little higher relevance for CPS systems, while the remaining categories draw a largely balanced picture.
Interestingly, \textit{dependability-related challenges}, which one would intuitively attribute to CPS microservices, played no major role in both domains, and received even slightly more mentions for enterprise application microservices.

\paragraph{Practices}
The comparison of practices shows a more balanced picture.
We again see a big gap related to \textit{organizational matters} and for \textit{architecture \& design} practices.
A closer look into the latter reveals that architectural solutions proposed for CPS account for distributed infrastructure, heterogeneity, and performance aspects.
Solutions for the enterprise domain, on the contrary, focus mainly on decomposing monolithic architectures into a suitable set of services, which correlates with the identified challenges in this category.
A similar difference shows up for the \textit{operations \& deployment} category.
Deployment solutions for CPS mainly address challenges of automation, placement of services, load balancing and monitoring.
Solutions for enterprise systems prioritize the automation aspect as well, but focus less on performance-related aspects than on integrated CI/CD pipelines that foster agility and short release cycles.
The inherently more challenging \textit{security} aspect for CPS systems is reflected by a higher number of practices around necessary access management.
\textit{Other aspects} include techniques and tools to improve autonomy or automation, managing product variants, and customer involvement.

\subsection{Comparing Rapid Review and Interview Results}
The primary studies in our rapid review reflect a strong focus on hardware-related aspects around the management and operation of heterogeneous environments, especially the identified challenges.
This is augmented with the complexity of big data processing and real-time conditions.
To address unsuitable CPS architectures that fail to address evolvability, operability, or performance efficiency, researchers proposed many concrete microservice-based architectures for various CPS contexts.
They predominantly offer solutions for efficient operations \& deployment in the cloud-edge continuum, but also for data processing \& modeling.

However, our interviewees from \companyName\ perceived most challenges with different importance.
The top two review categories (\textit{heterogeneity} and \textit{operations \& deployment}) played no major role for most analyzed systems, with only a few mentions.
Standardization in domains like healthcare (partly) addressed the former, while the latter was more relaxed because none of the analyzed systems needed dynamic edge-cloud deployment and scheduling at runtime.
Another reason for the lack of operational challenges might be that most systems were still early in the migration.
Instead, \textit{architecture \& design} was the dominant challenge category, with a focus on the difficulty of modularizing and decomposing monolithic systems.
Mentioned issues were commonly associated with legacy systems, e.g., huge size, hidden dependencies, or accumulated technical debt.
Consequently, the decomposition into services was a major challenge among our interviewees that was mostly described as a manual task with a lack of adequate tool support.

Another difference was the importance of \textit{team \& organizational challenges} in the interviews, while these seemed neglected in the review.
Establishing a DevOps culture and building developer confidence were seen as challenging, and the lack of employees experienced with current technologies could present a bottleneck.
Consequently, our interviewees reported compensatory practices around promoting team independence and collaboration, e.g., through conducting workshops.
Moreover, we also saw a preference for continuously evolving a system towards microservices and \mbox{DevOps} rather than a swift shift, which was, however, also due to strict quality and compliance requirements.

Contrary to the review, the interviews mostly reflected typical challenges and practices of enterprise microservices, as described by, e.g., Baškarada et al.~\cite{Baskarada2018} and Fritzsch et al.\cite{Fritzsch2019d}
Except for performance challenges (e.g., real-time processing requirements), typical CPS challenges like heterogeneity, dependability, or security were not seen as problematic for using microservices.
For several systems, that meant that not all microservices characteristics were desired for every part of the system.
However, this tendency has also been observed for enterprise applications.\cite{Bogner2019-ICSA,Zhang2019}
The good news about these findings is that many general approaches for microservices and DevOps are also valuable for the CPS domain.
Nonetheless, slight adaptations or tradeoffs are needed in some cases, as several CPS characteristics still seem to have an influence, albeit not a major one.

\subsection{Main Takeaways}
In the following paragraphs, we briefly synthesize the main takeaways of the combined study results and their implications.

\paragraph{The CPS and enterprise application literature differ in the importance of several microservice-related challenges.}
For example, \textit{heterogeneity} and \textit{performance efficiency} were perceived as much more challenging for CPS microservices than for enterprise application microservices.
Conversely, challenges related to \textit{team \& organization} were barely discussed in the CPS literature, while this was one of the top challenge categories for enterprise application microservices.
This may imply that the different usage scenarios and quality requirement prioritization between these two types of systems lead to a slightly different research focus.
Since the introduction of microservices in the CPS domain happened later than for enterprise applications, it might also mean that some industry-relevant challenges are not yet discussed frequently,
but will appear more in future publications, once the field matures a bit more.

\paragraph{Osmotic computing~\cite{Villari2016} was a prevalent microservices theme in the CPS literature, but not at Siemens.}
As a concrete example of the different research focus, many proposed CPS practices in the literature centered around the dynamic scheduling and deployment of microservices at runtime within the cloud-fog-edge continuum, e.g., based on the origin of the request or the current load of the system.
However, not a single one of the studied CPS at Siemens had this requirement.
It may be that osmotic computing is only relevant for CPS outside of Siemens's scope, e.g., very large, geographically distributed IoT environments, or that research in this space is still too early to broadly carry over to industry.

\paragraph{Improved development productivity and maintainability were the primary migration rationales at Siemens.}
CPS stakeholders predominantly wanted an architecture that would enable them to deliver features faster and more sustainably in the long run.
While this is also one of the main drivers in the enterprise application literature,\cite{Bogner2019-ICSA} other prominent microservices quality attributes like scalability, elasticity, or interoperability were rarely mentioned at Siemens.
A potential reason could be that the average CPS may not experience as frequent or rapid changes in the required throughput as typical internet-facing applications.

\paragraph{Typical CPS requirements like performance, reliability, security, or privacy were not perceived as major challenges regarding microservice migration at Siemens, but were mentioned for selected systems.}
Three systems named performance as a challenge with microservices (S1, S4, S5), two systems reliability (S1, S4), and one system each did the same for compliance (S3) and security (S6).
When explicitly queried for such quality attributes, other participants described that, while these requirements existed to a certain degree, the teams had experience in assuring these qualities, as they had to take care of them anyway in their domains.
Using microservices would not fundamentally change this, i.e., it would not be a migration-related challenge.
This implies that it is highly situational if assuring these quality attributes is challenging with CPS microservices and if it requires special methods or tool support.

\paragraph{Several studied CPS at Siemens did not target a full microservice architecture with all characteristics.}
Apart from S3, S4, and S6, the systems in our sample consciously decided to only aim for certain microservice characteristics and to leave out the rest, which was sometimes motivated by special CPS requirements.
For example, the \enquote{no shared databases} principle was consciously ignored for S4 because data consistency was just too important for their regulated domain.
This is important to keep in mind for researchers that work on methods or tool support for microservices in the CPS domain.

\paragraph{Most frequently mentioned challenges and practices at Siemens were not highly specific to CPS.}
Similar to the disconnect between the literature on CPS vs. enterprise application microservices, there are also several differences between the CPS literature and our case study findings at Siemens.
While the above sections mentioned several facets where the CPS domain had an influence on the usage of microservices, the overall sentiment regarding microservices challenges and practices was much closer to the enterprise application literature than to the CPS literature.
A potential reason could be that our rapid review contained many papers on large, decentralized IoT environments, which have some differences to the typical CPS at Siemens.
Additionally, it could be argued that a few of our studied systems are more platforms or applications from the CPS domain rather than actual CPS.
Nonetheless, it seems plausible that a large proportion of methods and tools for enterprise application microservices will at least be partly of value for CPS microservices.
Some flexibility for the pointed out differences will be helpful, but the extensiveness of the required adaptations should be subject to future research.

\subsection{Threats to Validity}
We generally aimed for a rigorous study design, but also made some tradeoffs regarding efficiency, as is common in rapid reviews.
The mixed-method approach and the collaboration between industry and academia could have partially mitigated this.
Still, several limitations need to be mentioned for our results, which we structure according to three common notions of validity described by Wohlin et al.\cite{Wohlin2012}

\textit{Construct validity} refers to a potential gap between operationalized measurements and the constructs targeted in the study, i.e., if the collected data is suitable for the intended analysis purpose.
In our study, data collection was text-based and qualitative, namely extractions from papers and interview transcripts with subsequent thematic analysis.
To combat potential threats in this area, we spent great care to arrive at a unified understanding of our central constructs in the research team, e.g., microservices, practices, challenges, rationales, etc.
Additionally, we introduced important constructs to interview participants, posed clarifying questions when encountering ambiguous terms, and sometimes briefly reflected our summarized understanding back to interviewees.
Nonetheless, there is a slight possibility that we interpreted a few instances differently than the authors of our primary studies or our participants.
Furthermore, interviews rely on the described experiences and opinions of participants, which are subjective.
A few interview statements could even be incorrect without us realizing.
In general, however, we estimate the impact of threats in this area to be very low.

\textit{Internal validity} is concerned with potential confounding factors and the rigor and consistency of the study design.
While we consciously chose a more lightweight research method for the literature review, we still followed a very systematic process.
At least two researchers decided if a study should be included or excluded.
While only a single researcher performed the data extraction for most papers, we only split up the work after having piloted the extraction sheet and improved our shared understanding of it with 10 papers.
Additionally, all important findings were discussed among several researchers to counteract potential subjective bias.
For the interviews, confidentiality and anonymity were provided to enable participants to talk freely about their experiences.
Participants also seemed not to be afraid to describe negative or more sensitive areas of their systems.
We also tried to limit our influence as interviewers on our participants, e.g., by avoiding leading questions or any judgmental statements.
However, despite these countermeasures, there is still a chance for non-mitigated influences on data collection, which we nonetheless perceive as non-critical.

Lastly, \textit{external validity} is concerned with how well the findings can be generalized to other settings or populations.
From the discussed facets of validity, this is the weakest one in our study.
A rapid review usually covers a lower percentage of the relevant literature, in our case via a stopping criterion (a maximum of 50 hits per search string and source).
We counteracted this via five different data sources and one round of snowballing, which should have led to sufficient study diversity.
Nonetheless, it is possible that the distributions for extractions could be different with a larger number of primary studies.
We also need to be careful to generalize the review findings to industry settings.
However, the more important limitation in this area is that we conducted a case study, a method that trades off external validity for result richness and depth.
Our interview-based findings at \companyName\ are valuable, but not all of them may be generalizable to other companies.
While Siemens is a global technology enterprise with diverse business units and projects in various domains, a certain degree of homogeneity between our units of analysis is to be expected.
Parts of our case study findings may therefore not be transferable to other CPS companies, especially to smaller or less experienced ones.
Even though we touched several CPS domains, our sample only contains seven systems that were described by nine CPS professionals.
Additional studies with other companies are therefore necessary to broaden the understanding of microservices and DevOps for CPS, which we try to enable by sharing our study artifacts for replications.

\section{Conclusion}
While challenges and practices regarding microservices and DevOps are well studied for enterprise applications, an overview of the cyber-physical systems domain is missing in this regard.
To address this, we formed an industry-academia collaboration, and conducted a rapid review (146 papers) followed by an interview-based case study with 9 different CPS experts at \companyName.
We found that, with a few exceptions, our interviewees' perception of challenges and their applied practices were much closer to the enterprise application domain than to the CPS literature.
Researchers and practitioners should take these findings into account when studying, developing, or modernizing CPS with microservices and DevOps.
We especially see potential for adaptation mechanisms in existing microservices methods and tools to support the mentioned challenges and peculiarities of CPS microservices.
Additionally, more research is needed to see to what degree our case study findings at Siemens are generalizable to other CPS companies.

\bibliography{references}

\end{document}